\documentclass[american,aps,manuscript]{revtex4}
\usepackage[T1]{fontenc}
\usepackage[latin1]{inputenc}
\usepackage{babel}
\usepackage{graphics}

\makeatletter

\providecommand{\LyX}{L\kern-.1667em\lower.25em\hbox{Y}\kern-.125emX\@}
\let\SF@@footnote\footnote
\def\footnote{\ifx\protect\@typeset@protect
    \expandafter\SF@@footnote
  \else
    \expandafter\SF@gobble@opt
  \fi
}
\expandafter\def\csname SF@gobble@opt \endcsname{\@ifnextchar[
  \SF@gobble@twobracket
  \@gobble
}
\edef\SF@gobble@opt{\noexpand\protect
  \expandafter\noexpand\csname SF@gobble@opt \endcsname}
\def\SF@gobble@twobracket[#1]#2{}

\usepackage[T1]{fontenc}
\usepackage[latin1]{inputenc}
\usepackage{babel}
\usepackage{graphics}



\makeatother
\begin{document}

\title{CdS/HgS/CdS Quantum Dot Quantum Wells: A Tight-binding Study}

\author{J. P\'{e}rez-Conde }

\address{Departamento de Física, Universidad Pública de Navarra E-31006, Pamplona,
Spain }

\author{A. K. Bhattacharjee }

\address{Laboratoire de Physique des Solides, UMR du CNRS, Université Paris-Sud,
F-91405, Orsay, France}

\begin{abstract}
We study the electronic properties of spherical quantum dot quantum
well nanocrystals within a symmetry-based tight-binding model. In
particular, the influence of a concentric monolayer of HgS embedded
in a spherical CdS nanocrystal of diameter \( 52.7 \)~\AA{} is analyzed
as a function of its distance from the center. The electron and hole
states around the energy gap show a strong localization in the HgS
well and the neighboring inner (core) interface region. Important
effects on the optical properties such as the absorption gap and the
fine structure of the exciton spectrum are also reported. 
\end{abstract}
\maketitle

\section{Introduction}

The interest in semiconductor nanocrystals (NC's) or quantum dots
(QD's) was first awakened by the size dependence of their electronic
properties which could allow, in principle, the manufacture of sensor,
lasers, etc. with specific features. One step further was the inclusion
of a layer of a different compound HgS in a CdS NC, for example, which
produces additional drastic changes in the optical properties in such
quantum dot quantum well (QDQW) NC's \cite{sm94,km99}. In fact, one
of the primary objectives of this manipulation was to avoid the, undesirable,
surface effects in the electronic states near the band gap \cite{mk96}.
In addition to the CdS/HgS/CdS case, systems based on ZnS/CdS \cite{le01}
have been studied.

On the theoretical side, a single band effective mass approximation
(EMA) model was first proposed \cite{sm94} followed by a multiband
EMA analysis \cite{jb98}. An atomistic theory was needed, however,
to adequately describe a QDQW where the well could be as thin as one
monolayer. Recently, a tight-binding (TB) model has been proposed
for CdS/HgS/CdS and ZnS/CdS/ZnS QDQW's \cite{le01,bj01}. This model
assumed, however, some simplifications: Instead of zincblende, fcc
structured NC's were studied and, also, the spin-orbit coupling was
neglected. Here we propose a symmetry-based TB theory which has been
previously used to account for the optical properties of CdSe \cite{pb01}
and CdTe \cite{pb99} NC's. Also, we take the zincblende crystalline
structure of the actual dots. We investigate the influence of the
HgS layer on the electronic and optical properties. In particular,
it is found that the charge distribution, gap and the fine structure
of the exciton spectrum are strongly dependent on the distance of
the monolayer from the NC center. Finally, we compare our results
with the available experimental data.

\begin{figure}
{\centering \resizebox*{1\columnwidth}{7cm}{\includegraphics{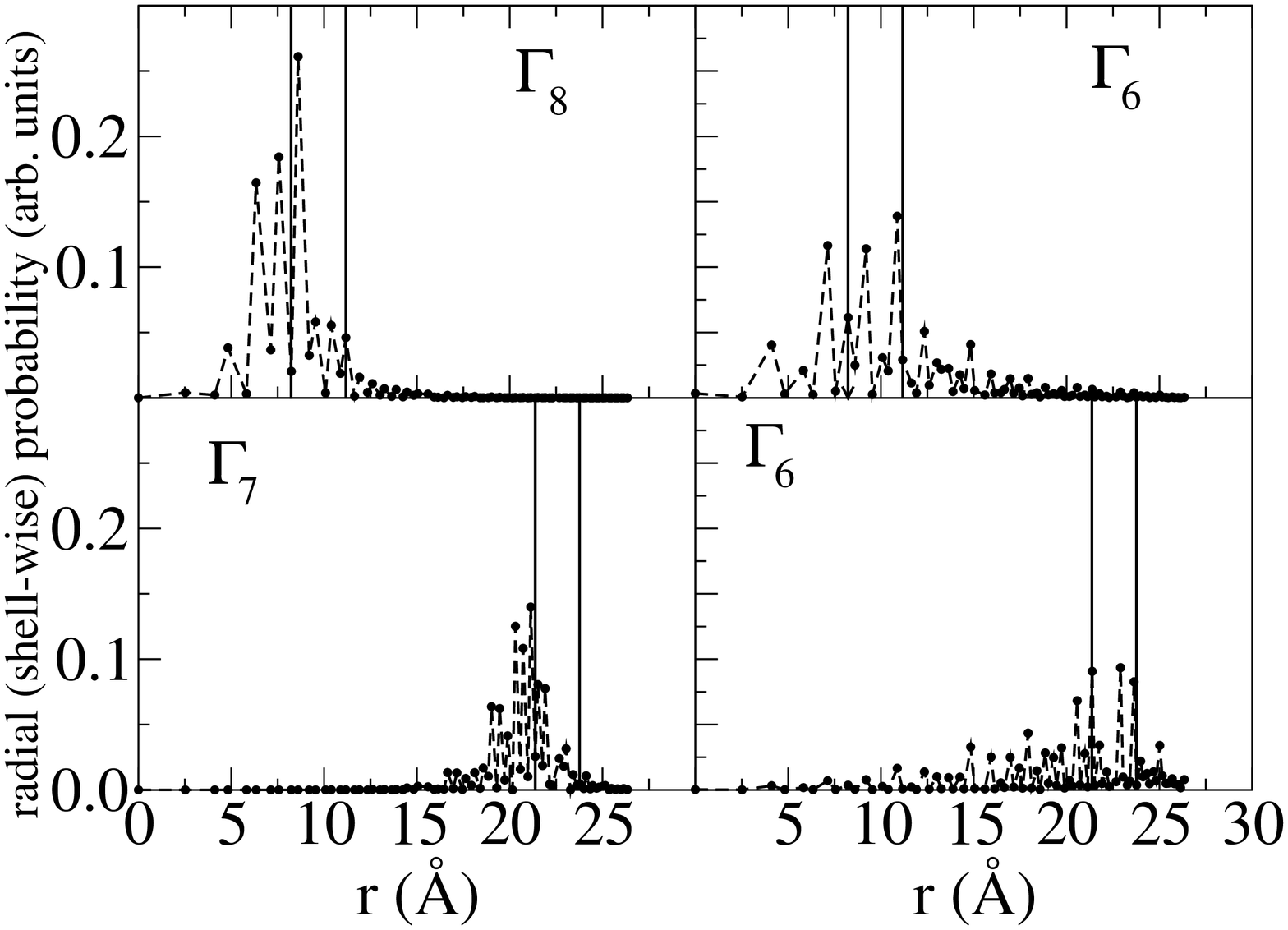}} \par}
\caption{Radial (shell-wise) probability of the HOMO (left) and LUMO (right).
The radius of the core, size of the well and clad are \protect\protect\( 8.2/3/15.1\protect \protect \) \AA{}
(top) and \protect\protect\( 21.4/2.4/2.5\protect \protect \) \AA{}
(bottom). Vertical bars indicate the HgS shell. The symmetry is also
indicated }
\label{fig1}
\end{figure}

\section{Theory}

The calculation of the one-particle TB Hamiltonian has been previously
described in detail \cite{pb99}. It is based on a semi-empirical
TB model for semiconductors \cite{ks82} which accounts for the bulk
properties. The interatomic hopping matrix elements are restricted
to the nearest neighbors. The CdS TB parameters used here are an adequate
modification of those proposed earlier \cite{ll89} to take now into
account the finite spin-orbit coupling. The HgS parameters have been
obtained from the CdS ones and account for the small bulk gap, \( 0.2 \)
eV, and the shift of the HgS valence band of \( 0.85 \) eV as in
Ref. \cite{br95}. We passivate the NC surface by placing a hydrogen
\( s \) orbital at each empty nearest-neighbor site on the surface
(dangling bond). We saturate the bonds so that the surface states
are several eV far from the gap edges. The NC's, of roughly spherical
shape, are constructed starting from a cation at the origin by successively
adding nearest-neighbor atoms through tetrahedral bonding. In this
work we study a NC with a fixed number of atoms, \( 3109 \), equivalent
to a diameter of \( 52.7 \) \AA{}. We use hereafter the word {}``shell''
to indicate the collection of atoms which are at the same distance
from the center. A chemical monolayer contains several shells and
their number depends on the layer radius. We reduce the Hamiltonian
to a block diagonal form by writing it in a symmetrized basis corresponding
to the double-valued representations \( \Gamma _{k} \) (\( k=6,7,8 \))
of \( T_{d} \).

When the Coulomb interaction is introduced the total Hamiltonian is
written in a many-body electron-hole basis. The details of the simplifying
approximations used to write the direct and exchange terms can be
found in Ref. \cite{pb01,lp98}. The absorption spectra are computed
following a simplified procedure as in \cite{pb01}. The dipole matrix
elements between different orbitals on the same atoms are taken from
\cite{fm81}. The fine structure of the lowest-energy transitions
is studied taking into account all the Hamiltonian terms, with as
many valence and conduction states as necessary to reach numerical
convergence.

\begin{table}
\begin{tabular}{|c|c|c|c|c|}
\hline 
\( r_{co}/r_{w}/r_{cl} \) (\AA) &
 Core &
 Well &
 Clad &
 electron radius (\AA) \\
\hline
\( 21.4/2.4/2.5 \)&
 0.47(0.81) &
 0.35(0.12) &
 0.15(0.06) &
 19.9 \\
\hline
\( 10.9/3.0/12.4 \)&
 0.23(0.17) &
 0.41(0.20) &
 0.33(0.62) &
 12.7 \\
\hline
\( 8.2/3.0/15.1 \)&
 0.19(0.09) &
 0.40(0.16) &
 0.41(0.75) &
 11.3  \\
\hline
\end{tabular}

\protect

\caption{The probability of presence of electron in the LUMO. The core radius
and also the well and clad sizes are given in the first column for
each case. In parenthesis we show the probability of presence in the
case of a simple CdS QD. The fourth column shows the electron radius,
to be compared with its value \protect\protect\( 15.5\protect \protect \) \AA{}
in a CdS QD}

\label{electron_size}
\end{table}

\begin{table}
\begin{tabular}{|c|c|c|c|c|}
\hline 
\( r_{co}/r_{w}/r_{cl} \) (\AA) &
 Core &
 Well &
 Clad &
 hole radius (\AA)\\
\hline
\( 21.4/2.4/2.5 \)&
 0.53(0.85) &
 0.44(0.11) &
 0.03(0.04) &
 20.7 \\
\hline
\( 10.9/3.0/12.4 \)&
 0.37(0.08) &
 0.55(0.21) &
 0.07(0.71) &
 10.5 \\
\hline
\( 8.2/3.0/15.1 \)&
 0.25(0.02) &
 0.63(0.10) &
 0.12(0.88) &
 8.5  \\
\hline
\end{tabular}

\protect

\caption{The probability of presence of hole in the HOMO. We use the same
notations and conventions as in Table \ref{electron_size}. The hole
radius in a CdS QD is \protect\protect\( 18.5\protect \protect \) \AA{} }

\label{hole_size}
\end{table}

\section{Results}

In Fig.~\ref{fig1} we show the radial probability of the highest
occupied molecular orbital (HOMO) and the lowest unoccupied molecular
orbital (LUMO) for two different situations. When the HgS monolayer
is included, both the electron and hole show an enhancement of their
presence within the well. From Table~\ref{electron_size} and Table~\ref{hole_size}
it can be also seen that the hole is always more localized than the
electron. Moreover, the wells of smaller radius are more efficient
for trapping the particles. Finally, a careful inspection of Fig.~\ref{fig1}
shows an increase of the hole density on the inner part of the interface
neighborhood. The states nearest in energy to the HOMO and the LUMO
show similar shapes.

\begin{table}
\begin{tabular}{|c|c|c|c|}
\hline 
\( D=52.7 \) \AA{}&
 HOMO (eV) &
 LUMO (eV) &
 Gap (eV)\\
\hline
\( 26.3/0.0/0.0 \)&
 -0.093(8) &
 2.870(6) &
 2.664 \\
\hline
\( 21.4/2.4/2.5 \)&
 0.284(7) &
 2.321(6) &
 1.936 (1.953)\\
\hline
\( 10.9/3.0/12.4 \)&
 0.333(8) &
 2.187(6) &
 1.696(1.768) \\
\hline
\( 8.2/3.0/15.1 \)&
 0.303(8) &
 2.285(6) &
 1.798(1.851)  \\
\hline
\end{tabular}

\protect

\caption{The highest (lowest) occupied (unoccupied) valence (conduction) levels
in the second (third) column along with the symmetry in parenthesis.
The results correspond to a dot of \protect\protect\( 52.7\protect \protect \) \AA{}
of diameter. The same cases as in previous tables are considered.
The optical gap is given in the last column. We give also the energy
of the next important peak in parenthesis (see also Fig. \ref{fig2})}

\label{gaps}
\end{table}

\begin{figure}
{\centering \resizebox*{1\columnwidth}{7cm}{\includegraphics{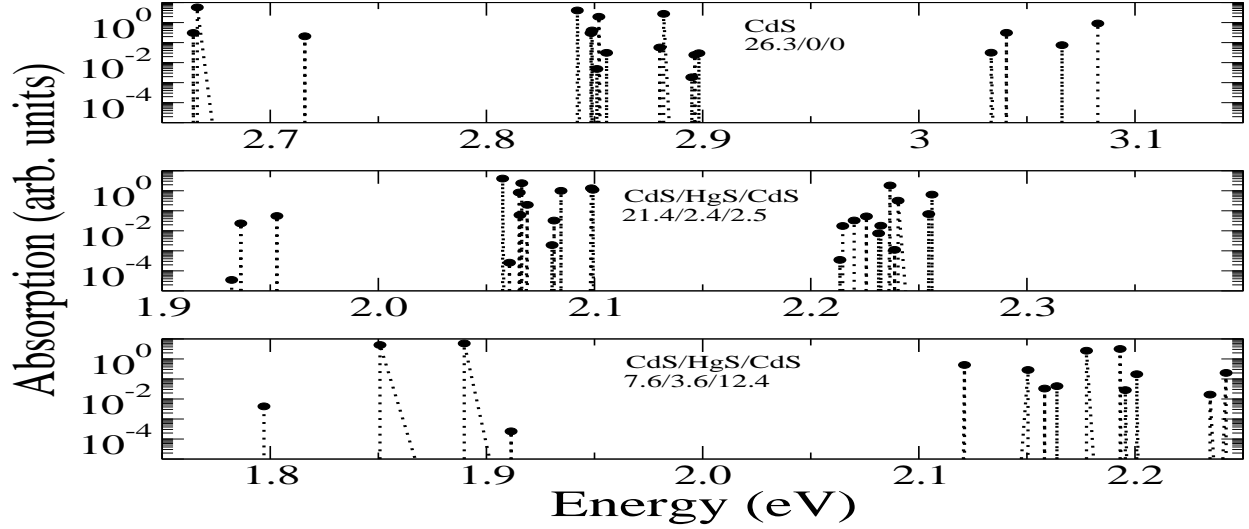}} \par}

\caption{Fine structure of the optical absorption spectrum for three different
NC's. We show the same window of energy, \protect\protect\( 0.5\protect \protect \)
eV, for each case. The dotted lines help distinguish the different
transitions }

\label{fig2}
\end{figure}

In Table \ref{gaps} we show the values of the HOMO and LUMO energies
and the optical gap obtained from the energy of the first allowed
exciton state. The overall trend indicates that the gap decreases
as the monolayer radius decreases. The gap of the \( 52.7 \) \AA{}
wide CdS dot, \( 2.666 \) eV, is in good agreement with the value
measured by Schoss \textsl{et al.} \cite{sm94}, \( 2.62 \) eV. The
structure of the lowest energy optical spectra shows also significant
differences between the CdS QD and the CdS/HgS/CdS QDQW. In Fig. \ref{fig2}
the absorption spectra of three different dots are shown. The CdS
QD presents the first peak at \( 2.664 \) eV with a relative intensity
of \( 0.303 \). Very close in energy there is a more important peak
at \( 2.666 \) eV which shows an intensity of \( 5.753 \). When
a HgS monolayer is included the first transition (exciton ground state)
gets relatively weaker as the HgS monolayer radius decreases. When
the HgS layer radius is \( 21.4 \) \AA{} the intensities of the exciton
ground state and the states close in energy are smaller than in the
CdS QD. As the radius decreases further, at \( 7.6 \) \AA, there
is only a {}``dark'' exciton ground state separated \( 53 \) meV
from the {}``bright'' states. The agreement with previous EMA \cite{sm94,jb98}
and TB \cite{bj01} calculations is good for the one-electron description:
The energies and charge localization. As for the exciton, a comparison
is difficult because we take the full many-body Hamiltonian and, also,
the QD sizes studied here are smaller.

\section{Conclusion}

We have presented a TB model adequate to describe NC heterostructures.
We have considered the inclusion of a single QW monolayer of HgS in
a spherical CdS NC. The HOMO and LUMO states near the gap edges are
localized in the QW layer and its proximity. The effects on the optical
properties are also important. In particular, the absorption gap decreases
when the layer radius decreases. The absorption spectrum also changes:
The relative intensity associated with the {}``dark'' exciton state
is a decreasing function of the layer radius.

\end{document}